\documentclass[showpacs,prb,twocolumn,groupedaddress,floatfix]{revtex4-1}

\usepackage{bm}
\usepackage{pstricks}
\usepackage{amsmath,amssymb,amsbsy}
\usepackage{graphicx,subfigure}
\usepackage{verbatim}
\usepackage[utf8]{inputenc}
\usepackage[USenglish]{babel}
\usepackage[T1]{fontenc}
\usepackage{dcolumn}
\usepackage{multirow}
\usepackage{latexsym}
\usepackage{tikz}
\usepackage{pgf}
\usepackage{pbox}
\usepackage{float}
\usepackage{multirow}
\usepackage{braket}
\usepackage{booktabs}
\usepackage{supertabular}
\usepackage{hyperref}

\usepackage{xcolor}
\hypersetup{
    colorlinks,
    linkcolor={blue!50!black},
    citecolor={blue!50!black},
    urlcolor={blue!80!black}
}

\newcommand{\be}{\begin{equation}}
\newcommand{\ee}{\end{equation}}
\newcommand{\bea}{\begin{eqnarray}}
\newcommand{\eea}{\end{eqnarray}}

\newcommand{\im}{\mathrm{i}}
\newcommand{\e}{\mathrm{e}}

\begin{document}

\title{Excitons in hexagonal boron nitride single-layer: a new platform for polaritonics in the ultraviolet}

\author{F. Ferreira}

\affiliation{Centro de F\'{i}sica and Departamento de F\'{i}sica, Universidade
do Minho, Campus de Gualtar, Braga 4710-057, Portugal}

\author{A.~J. Chaves}

\affiliation{Centro de F\'{i}sica and Departamento de F\'{i}sica, Universidade
do Minho, Campus de Gualtar, Braga 4710-057, Portugal}

\author{N.~M.~R. Peres}

\affiliation{Centro de F\'{i}sica and Departamento de F\'{i}sica and QuantaLab, Universidade
do Minho, Campus de Gualtar, Braga 4710-057, Portugal}
\affiliation{International Iberian Nanotechnology Laboratory (INL), Av. Mestre José Veiga, 4715-330 Braga, Portugal}

\author{R.~M. Ribeiro}

\affiliation{Centro de F\'{i}sica and Departamento de F\'{i}sica and QuantaLab, Universidade
do Minho, Campus de Gualtar, Braga 4710-057, Portugal}
\affiliation{International Iberian Nanotechnology Laboratory (INL), Av. Mestre José Veiga, 4715-330 Braga, Portugal}

\date{\today}


\begin{abstract}

The electronic and optical properties of 2D hexagonal boron nitride are studied using first principle calculations.
GW and BSE methods are employed in order to predict with better accuracy the excited and excitonic properties of this material.
We determine the values of the band gap, optical gap, excitonic binding energies and analyse the excitonic wave functions.
We also calculate the exciton energies following an equation of motion formalism and the Elliot formula, and find a very good agreement
with the $GW$+BSE method. The optical properties are studied for both the TM and TE modes, showing that 2D hBN is a good candidate
to polaritonics in the UV range. In particular it is shown that
a single layer of h-BN can act as an almost perfect mirror for
ultraviolet electromagnetic radiation.

\end{abstract}

\maketitle
\section{Introduction}

Two dimensional hexagonal boron nitride (hBN), also called by some white graphene, is an electrical insulator in which the boron (B) and nitrogen (N) atoms
are arranged in a honeycomb lattice and are bounded by strong covalent bonds.
Like graphene, hBN has good mechanical properties\cite{Bao2016} and high thermal conductivity.\cite{0022-3727-49-26-265501}
Specially interesting is the possibility of using hBN as a buffer layer in van der Waals heterostructures, namelly ones comprised by layers of h-BN/graphene.\cite{PhysRevB.93.235403}
 Hexagonal boron nitride layer can serve as a dielectric or a substrate material for graphene in order to improve its mobility\cite{doi:10.1021/acs.nanolett.5b04840}
and open a gap\cite{Jung1}.
It can also be used to improve the thermoelectric performance of graphene.\cite{Duan13122016}

Yet, its electronic properties differ significantly from graphene.
Graphene $\pi$ and  $\pi*$ electronic bands have a linear dispersion at the K point, whereas in hBN there is a lift of the degeneracy at the same point
and a wide band gap greater than 7 eV is formed, at least within an independent electron picture.
That would, in principle, make it ideal for optoelectronic devices in the deep ultraviolet region \cite{doi:10.1021/acs.cgd.6b00398, 2053-1583-4-2-021023}.
As we will see, however, excitonic effects play an important role in this material:
excitonic peaks are created at the near UV, and this is a much more useful electromagnetic spectral range, when compared to the deep UV.

The optical properties of monolayer hBN at the UV range are characterized by the exciton with a corresponding
optical band gap calculated in the range $5.30$--$6.30$ eV (see Sec. \ref{sec:GW}). The presence of
the exciton in this range can be used to excite exciton-polaritons,
that share some properties with surface plasmon-polaritons \cite{basov2016polaritons,low2017polaritons}.
 Therefore, the UV optical properties of hBN can be used as an alternative to the emerging field of UV plasmonics.
\cite{watanabe2011deep,mattiucci2012ultraviolet,mcmahon2013plasmonics,yang2013ultraviolet,maidecchi2013deep,ross2014aluminum,
watson2015rhodium,alcaraz2015rhodium,gutierrez2016oxide,gutierrez2018plasmonics}
The plasmonics in UV range also attracts interest in biological tissue \cite{kumamoto2011deep} as consequence of the
resonances in nucleotide bases and aromatic amino acids. Plasmonics in this ranges relies in poor metals  \cite{knight2012aluminum,maidecchi2013deep,ross2014aluminum, mcmahon2013plasmonics} and
Rhodium \cite{watson2015rhodium,alcaraz2015rhodium,gutierrez2018plasmonics}.

Because of the difficulty of its synthesis, few experimental works have been done for hBN single layer.
Also, to study and probe its electronic and optical properties it is necessary to work in UV range.
To our best knowledge only one experimental work\cite{Nagashima1995} has been produced that studies the electronic properties of 2D hBN.
Those authors observed the band structure of BN monolayer on Ni(111) surface by using angle-resolved ultraviolet-photoelectron spectroscopy and angle-resolved
secondary-electron-emission spectroscopy.
Because the bond between the interface of h-BN and Ni(111) is weak, the band-structure observed can be regarded as that of the monolayer h-BN.
The band gap was determined to be $\sim$ 7 eV and after a comparison with theoretical works, the authors conclude that the band gap is estimated
to be within the range of 4.6 to 7.0 eV, too wide when compared with numerical results.
These theoretical works were based on first principles calculations using Density Functional Theory (DFT).
It is well known that DFT does not predict with good accuracy the electronic and optical properties of semiconductors and insulators.
Accurate values require a theory that include many-body effects like the $GW$ approximation.\cite{PhysRev.139.A796,HedinS}
To obtain optical properties, a theory that includes the excitonic properties is also needed.
Usually, the Bethe-Salpeter equation (BSE)\cite{PhysRev.84.1232, PhysRevLett.80.4510} is used.

There are several works in the literature that used the $GW$ approximation\cite{PhysRevB.51.6868, PhysRevB.80.155453, Aaltodoc}
and $GW$+BSE\cite{PhysRevLett.116.066803, PhysRevLett.96.126104, PhysRevB.94.125303} on 2D h-BN.
The results from these works vary significantly, as can be seen in Table \ref{table:GW_summary_P}.
There is no agreement even on whether the gap is direct or indirect.
Convergence can be an issue in $GW$ and BSE calculations as can be seen in References \onlinecite{PhysRevB.96.115431} and \onlinecite{PhysRevLett.105.146401}.
It is likely that the works summarized in Table \ref{table:GW_summary_P} use different criteria for convergence and that may explain the differences.

\begin{table}[b]
\centering
\caption
{Several band gaps calculated in this work and by other authors using $GW_0$, $G_0W_0$ and BSE, in eV.
K$\rightarrow$K indicates the direct gap at the K point, K$\rightarrow\Gamma$ indicates the indirect gap from the K to the $\Gamma$ point,
O. Gap is the optical gap and EBE the excitonic binding energy.}
\begin{ruledtabular}
\begin{tabular*}{\columnwidth}{@{\extracolsep{\fill} }lccccc}
Reference                                    & Calculation     & K$\rightarrow$K  & K$\rightarrow\Gamma$  & O. Gap  & EBE   \\ \midrule
   This work                                 & $G_0W_0$ + BSE  & 7.77 & 7.32 & 5.58 & 2.19\\
   Ref. \onlinecite{PhysRevLett.96.126104}   & $GW_0$ + BSE    & 7.80 & -    & 6.30 & 2.10 or 1.50 \\
   Ref. \onlinecite{PhysRevLett.116.066803}  & $G_0W_0$ + BSE  & 7.36 & -    & 5.30 & 2.06\\
   Ref. \onlinecite{PhysRevB.94.125303}      & $G_0W_0$ + BSE  & 7.25 & -    & -    & 1.90 \\
   Ref. \onlinecite{Aaltodoc}                & $G_0W_0$        & -    & 7.40 & -    & - \\
   Ref. \onlinecite{PhysRevB.80.155453}      & $GW_0$          & -    & 6.86 & -    & - \\
   Ref. \onlinecite{PhysRevB.51.6868}        & $G_0W_0$        & -    & 6.00 & -    & - \\
 \end{tabular*}
\end{ruledtabular}
\label{table:GW_summary_P}
\end{table}

A small number of bands used in the calculation\cite{PhysRevB.51.6868,PhysRevB.80.155453} or not using a truncation to avoid interaction
with periodic images\cite{Aaltodoc} may also explain some differences.
Sometimes there is some ambiguity between the value stated for the gap and the one that can be obtained from the absorption spectrum presented.\cite{PhysRevLett.96.126104}
More difficult to explain are the values obtained in Ref. \onlinecite{PhysRevB.94.125303}.
They differ significantly from our work and others, although they seem to have converged the calculations carefully.
One explanation may be that they fixed the lattice constant at the experimental value, instead of relaxing the unit cell.
The experimental lattice constant may not match the value that actually optimizes the system and can influence the values of the gaps in the electronic band-structure.
An effective-energy technique\cite{PhysRevB.82.041103} was adopted in  Ref. \onlinecite{PhysRevLett.116.066803}.
That technique allowed the calculation of the screened Coulomb interaction $W$ to be converged with only 90 bands and 60 bands for the self-energy $\Sigma$ calculation.
The use of such technique certainly will produce some differences in the final results.

In this work we   clarify whether the gap in hBN is direct or indirect, as well as their values and the exciton energies.
We also calculate the excitonic spectra using an equation of motion formalism and the Elliot formula, fitting it with the $GW$+BSE calculations
thus obtaining a validation of the method.
In Section \ref{sec:GW} we describe the details of the $G_0W_0$ calculations and results.
In Section \ref{sec:BSE} we show the results of the BSE calculations.
Both $G_0W_0$ and BSE calculations were performed with the software package {\sc BerkeleyGW}.\cite{Deslippe20121269,Louie1986,Louie2000}
Section \ref{sec:Elliot} presents the equation of motion formalism and the results for the excitonic properties of monolayer hBN.
In Section \ref{sec:Ex_pol} we study the properties of exciton-polaritons of hBN and we show that a monolayer of  hBN can be used as a UV mirror.
We finally draw the conclusions in Section \ref{sec:conclusion}.

\section{$G_0W_0$ results}
\label{sec:GW}

$G_0W_0$ calculations were done on top of DFT calculations with a scalar-relativistic norm-conserving pseudopotential.
The software package {\sc Quantum ESPRESSO}\cite{QE-2009} was used for the DFT calculations.
The details of the  DFT calculations are summarized in Table \ref{DFT_details}.
For $G_0W_0$ calculations, a truncation technique is needed due to the non-local nature of this theory.

\begin{table}[b]
\centering
\caption{Details of DFT calculations.}
\begin{ruledtabular}
 \begin{tabular}{@{}lr@{}}
Exchange-correlation functional       &    GGA-PBE~\cite{PhysRevLett.77.3865}    \\
Plane-wave cut-off                    &   70 Ry              \\
$\textbf{K}$-point sampling (Monkhorst-Pack)\cite{PhysRevB.13.5188}  &   $6\times 6\times 1$   \\
Interlayer distance &   15.8 \AA       \\
Lattice constant    &   2.5 \AA       \\
\end{tabular}
\end{ruledtabular}
\label{DFT_details}
\end{table}

\begin{figure}[t]
\centering
\hspace*{-0.3cm}
\includegraphics[scale=0.34]{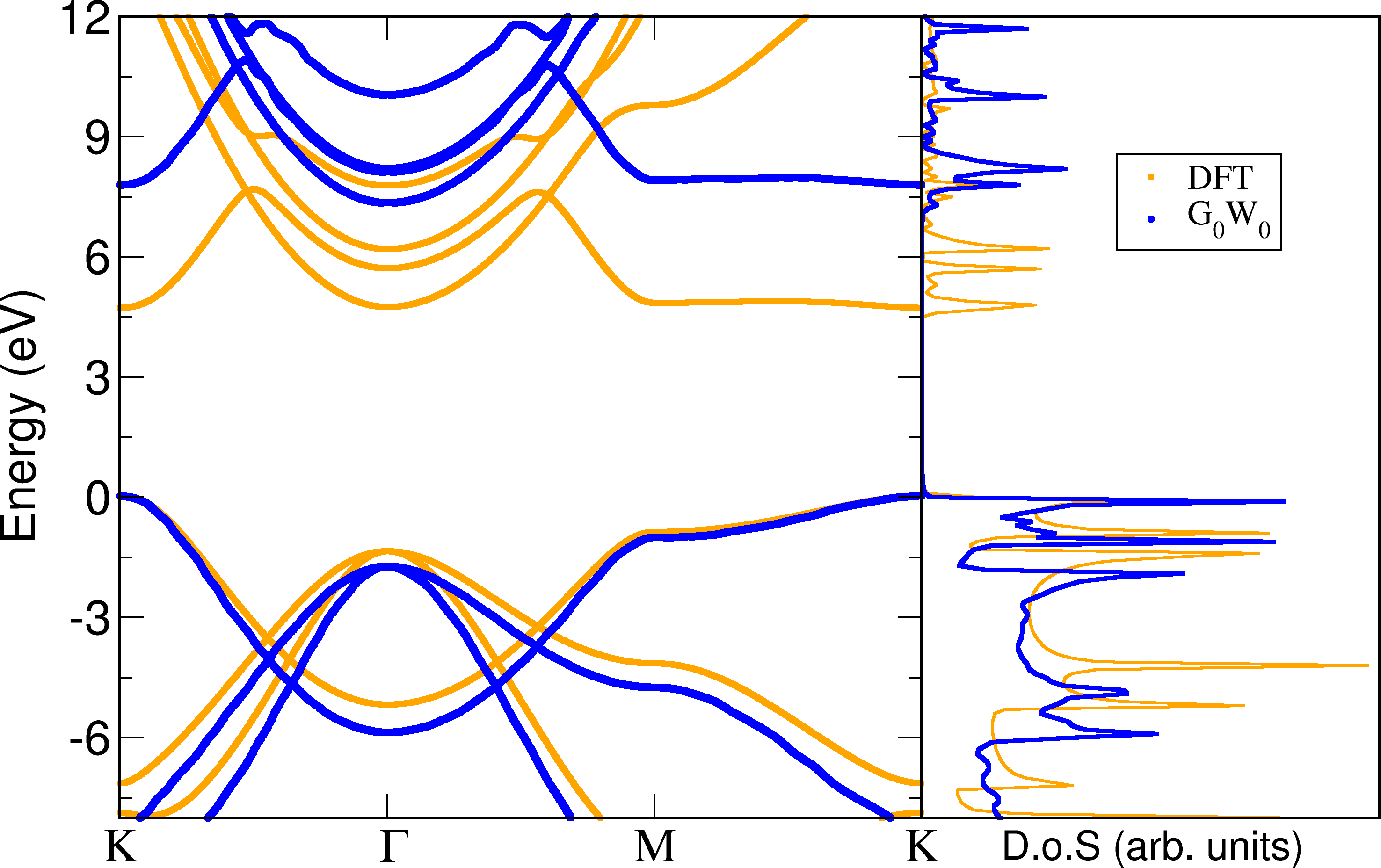}
\caption{(Color online) Electronic band structure (left) and electronic density of states of h-BN (right) for both DFT and $GW$ calculations. }
\label{BSE_GW16_banddosw_final}
\end{figure}

We found that for DFT calculations a grid of $6\times 6\times 1$ $\textbf{k}$-points is enough to reach convergence.
For the $GW$ calculations, a grid of $16 \times 16 \times 1$ $\mathbf{k}$-points and
a cut off energy of 22.6 Ry and 1100 bands were needed for the dielectric matrix calculations.
For the $\Sigma$ self-energy calculation we used a cut off energy of 22.6 Ry and 1000 bands.
The results obtained for the electronic band gap are summarized in Table \ref{gap_results_final}.
They show that a monolayer of hBN is a wide band-gap indirect-gap material.
Fig. \ref{BSE_GW16_banddosw_final} presents the electronic band structure and electronic density of states for both DFT and $GW$ calculations.

\begin{table}[t]
\centering
\caption{$G_0W_0$ gap values for the transitions K$\rightarrow \Gamma$, K$\rightarrow$K  and  $\Gamma\rightarrow\Gamma$,
and optical gap and exciton binding energy (EBE) obtained from BSE in this work. The results show that hBN is an indirect-gap insulator.}
\begin{ruledtabular}
 \begin{tabular}{@{}lccccr@{}}
Transition  &   K$\rightarrow \Gamma$  & K$\rightarrow$K  &  $\Gamma\rightarrow\Gamma$  & Optical & EBE  \\ \midrule
Energy [eV]            &   7.32                        &     7.77             &   9.07  & 5.58    & 2.19  \\
\end{tabular}
\end{ruledtabular}
\label{gap_results_final}
\end{table}

As mentioned in the Introduction, the only experimental work we are aware off is the one from Ref. \onlinecite{Nagashima1995},
which in fact estimates the band gap based on theoretical works that used mean field calculations to predict the electronic properties of bulk h-BN.
Mean field theories such as DFT underestimate the band gap value of semiconductors and insulator materials.
They obtain a wide range of possible values, from 4.6 to 7.0 eV.
We believe that a value closer to 7.0 eV is more reliable, since the gap value for the bulk materials are lower when compared to the monolayer counterpart.
And is actually closer to the ones obtained by works referred in Table \ref{table:GW_summary_P}.
Still, more experimental work is needed.

Ref. \onlinecite{Nagashima1995} also calculated the width of the valence bands,
and they found no good agreement with theoretical works of the time.
Table \ref{tab:width_results} shows the width of the valence bands as calculated with DFT, $GW$
and the experimental determination of Ref. \onlinecite{Nagashima1995}.
The $\pi$-band is the one that has its highest energy at the K-point,
while the $\sigma_1$ and $\sigma_2$ are the bands that have the highest energy at the $\Gamma$ point.
Table \ref{tab:width_results} shows that  DFT results differ from the experimental ones by values greater than 0.5 eV in all cases.
On the other hand, $G_0W_0$ results differ from the experimental results by values equal or smaller than 0.1 eV.

\begin{table}[b]
\centering
\caption{Width of the valence bands. $\pi$ band is the top valence band at K point.
$\sigma_1$ and $\sigma_2$ are the bands that have the highest energy at the $\Gamma$ point.
The difference is in the width of the bands which is greater for $\sigma_2$. }
\begin{ruledtabular}
 \begin{tabular}{@{}lccc@{}}
Width of bands [eV]             & $\pi$ band & $\sigma_1$ band &  $\sigma_2$ band    \\
\midrule
This work (DFT)                 & 5.20       &    5.78         &   7.49      \\
This work ($G_0W_0$)            & 5.90       &    6.42         &   8.24  \\
Experimental\cite{Nagashima1995}& 5.80       &    6.50         &   8.20   \\
\end{tabular}
\end{ruledtabular}
\label{tab:width_results}
\end{table}

We also calculated the effective masses of the highest valence band and lowest conduction band using both $G_0W_0$ and DFT (Table \ref{tab:effec_masses}).
We found no differences between $G_0W_0$ and DFT, except for the effective mass at K$\rightarrow {\Gamma}$ on the first conduction band
(DFT value greater by $0.08 m_e$).
Thus we conclude that DFT calculations are reliable to obtain the values of the effective masses in this material.

\begin{table}[t]
\centering
\caption{Effective masses (in electron mass ($m_e$) units) for h-BN calculated using $G_0W_0$.
The arrow indicate the direction in which the effective mass is calculated.}
\begin{ruledtabular}
 \begin{tabular}{@{}lcccc@{}}
\multicolumn{5}{c}{Effective mass $m^*/m_e$} \\ \midrule
Symmetry points & K$\rightarrow_{\Gamma}$ & $\Gamma \rightarrow_{\mathrm{K}}$ & $\Gamma \rightarrow_{\mathrm{M}}$ & M$\rightarrow_{\Gamma}$ \\
\midrule
Valence band    & 0.63  & 0.82  & 1.09  & 0.46   \\
Conduction band & 0.83  & 0.95  & 1.27  & 0.35  \\
\end{tabular}
\end{ruledtabular}
\label{tab:effec_masses}
\end{table}

Ref. \onlinecite{PhysRevB.51.6868} also calculated the effective mass at the $\Gamma$ point for the conduction band,
and obtained a value of $(0.95 \pm 0.05)m_e $ with only slight variations for different planar directions.
In our work we obtained differences of $0.3 m_e$ between different directions in reciprocal space at the $\Gamma$ point.

\section{BSE results}
\label{sec:BSE}

After determining the conduction and valence band states, the electron-hole pair states are determined using the Bethe-Salpeter (BSE) equation.
The imaginary part of the dielectric function $\epsilon_2(\omega)$ is then\cite{Louie2000}
\begin{equation}
 \epsilon_2(\omega) = \dfrac{8 \pi^2 e^2}{\omega^2}\sum_S \left| \mathbf{e}  \cdot \bra{0}\mathbf{v}\ket{S}\right|^2 \delta(\omega - \Omega^S)
\label{eq:epsilon_im_BSE}
\end{equation}
where  $\Omega^S$ is the  energy for an  excitonic state $S$, $\bra{0}\mathbf{v}\ket{S}$ is
the velocity matrix element, and $\mathbf{e}$ is the direction of the polarization of incident light with energy $\omega$.
$e$ is the electron charge.

If we do not consider excitonic effects, the expression becomes a transition between single particle states\cite{Louie2000}
\begin{equation}
 \epsilon_2(\omega) = \frac{8 \pi^2 e^2}{\omega^2} \sum_{vc\mathbf{k}}
\left| \mathbf{e}  \cdot \bra{v\mathbf{k}}\mathbf{v}\ket{c\mathbf{k}}\right|^2
\delta(\omega - \left( E_{c\mathbf{k}} - E_{v\mathbf{k}}\right) )
\label{eq:epsilon_im_RPA}
\end{equation}
which is a random phase approximation (RPA). The labels
$v$ $(c)$ denotes valence (conduction) band states, and $\mathbf{k}$ denotes the single particle momentum (only vertical transitions are considered).

\begin{figure}[h]
\centering
\includegraphics[scale=0.31]{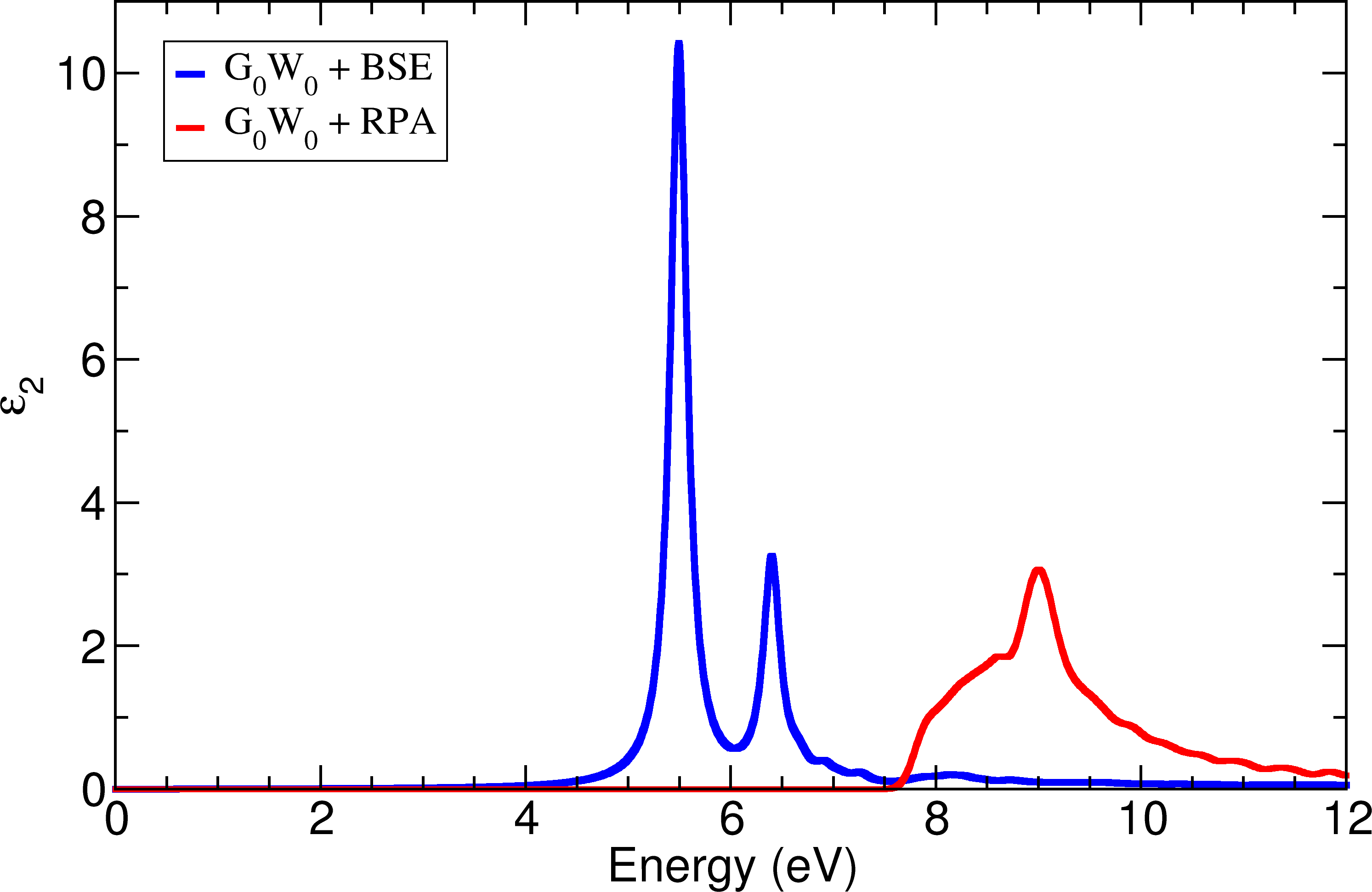}
\caption{(Color online) Imaginary part of dielectric function of 2D h-BN. The blue (red) lines represent the BSE (RPA) imaginary part of dielectric function.}
\label{BSE_GW16_final}
\end{figure}

\begin{figure}[t]
\centering
\includegraphics[scale=0.31]{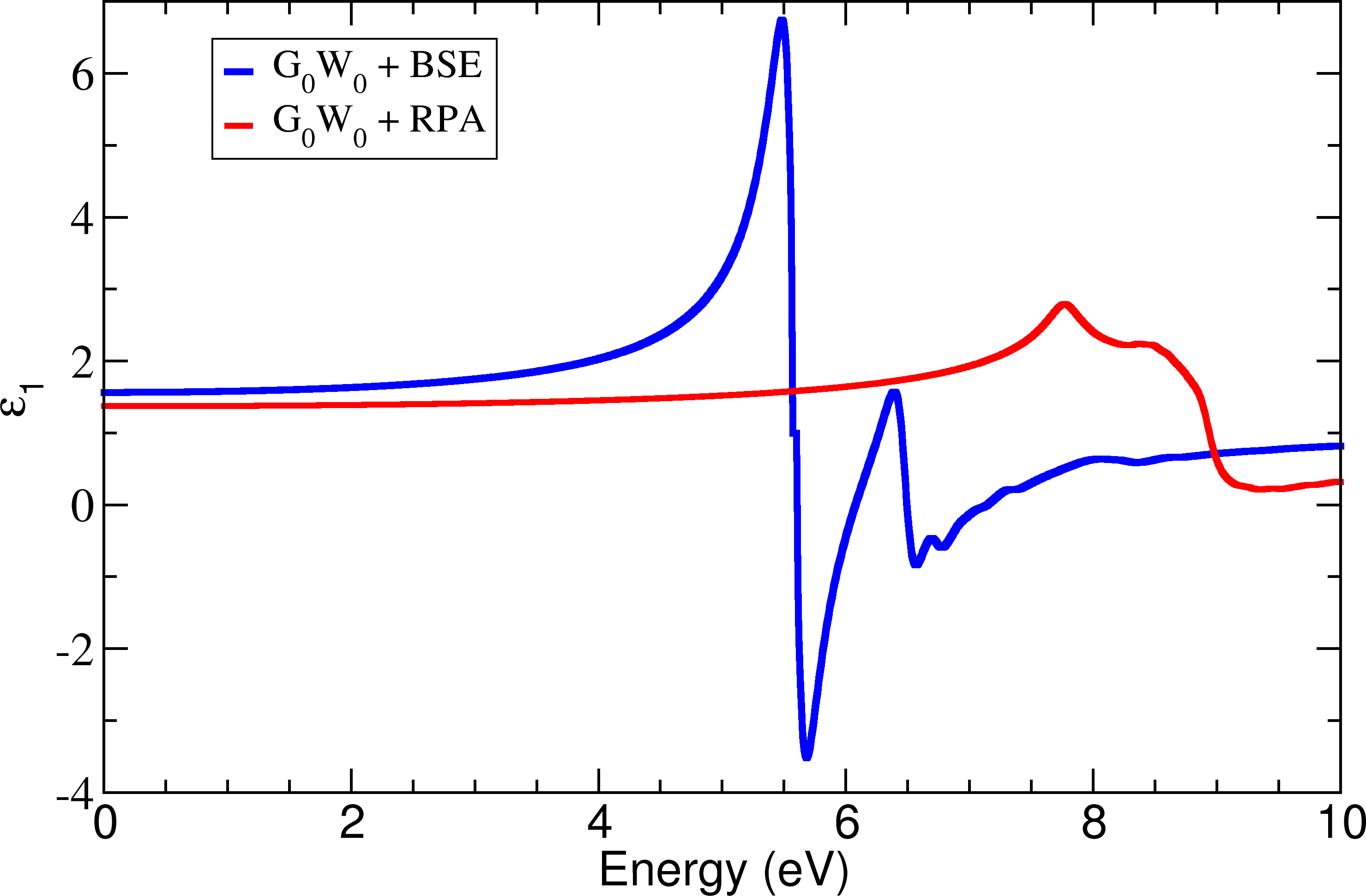}
\caption{(Color online) Real part of the dielectric function of 2D h-BN. The blue (red) lines represent the BSE (RPA) real part of dielectric function.}
\label{fig:real_part_of_BSE}
\end{figure}

Fig. \ref{BSE_GW16_final} shows the imaginary part of the dielectric function calculated by BSE,
done on top of a $G_0W_0$ calculation with a grid of $16 \times 16 \times 1$ $\mathbf{k}$-points.
The convergence of the $G_0W_0$ band structure with a particular grid of $\mathbf{k}$-points does not imply that BSE will be converged with the same grid.
An interpolation with a fine grid of $120 \times 120 \times 1$  $\mathbf{k}$-points was needed to achieve convergence.
Fig. \ref{BSE_GW16_final} also shows the imaginary part of dielectric function without excitonic effects.
The first peak has an energy of 5.58 eV and the second peak has an energy of 6.48 eV.
In Table \ref{gap_results_final} we summarize the gap values of the band structure, the optical gap and the excitonic binding energy.
Fig. \ref{fig:real_part_of_BSE} shows the real part of the dielectric function  calculated with and without excitonic effects.

\begin{figure}[h]
\centering
\includegraphics[scale=0.3]{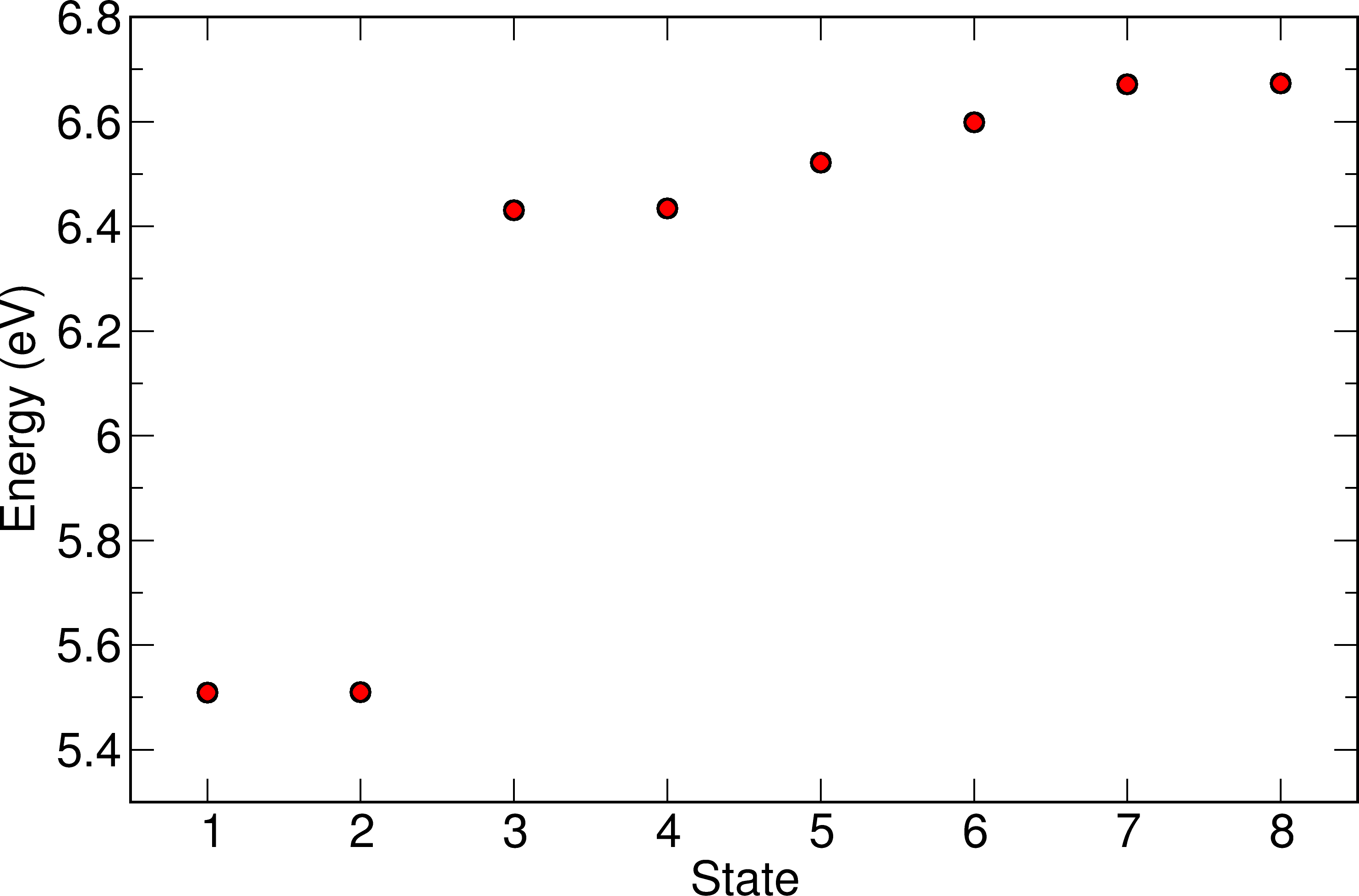}
\caption{(Color online) Excitonic energies for the lowest energy exciton states.
The system has a $C_{3v}$ symmetry with three representations: $A_1$, $E$ and $A_2$.
The states 1 to 4 have $E$ symmetry and are valley degenerate;
states 5 and 6 have $A_2$ and $A_1$ symmetries respectively and are non degenerate (see Ref. \onlinecite{PhysRevB.94.125303}).}
\label{EX_energies}
\end{figure}

\begin{figure*}[h]
 \centering
\begin{tabular}{cccc}
 1\includegraphics[scale=0.1]{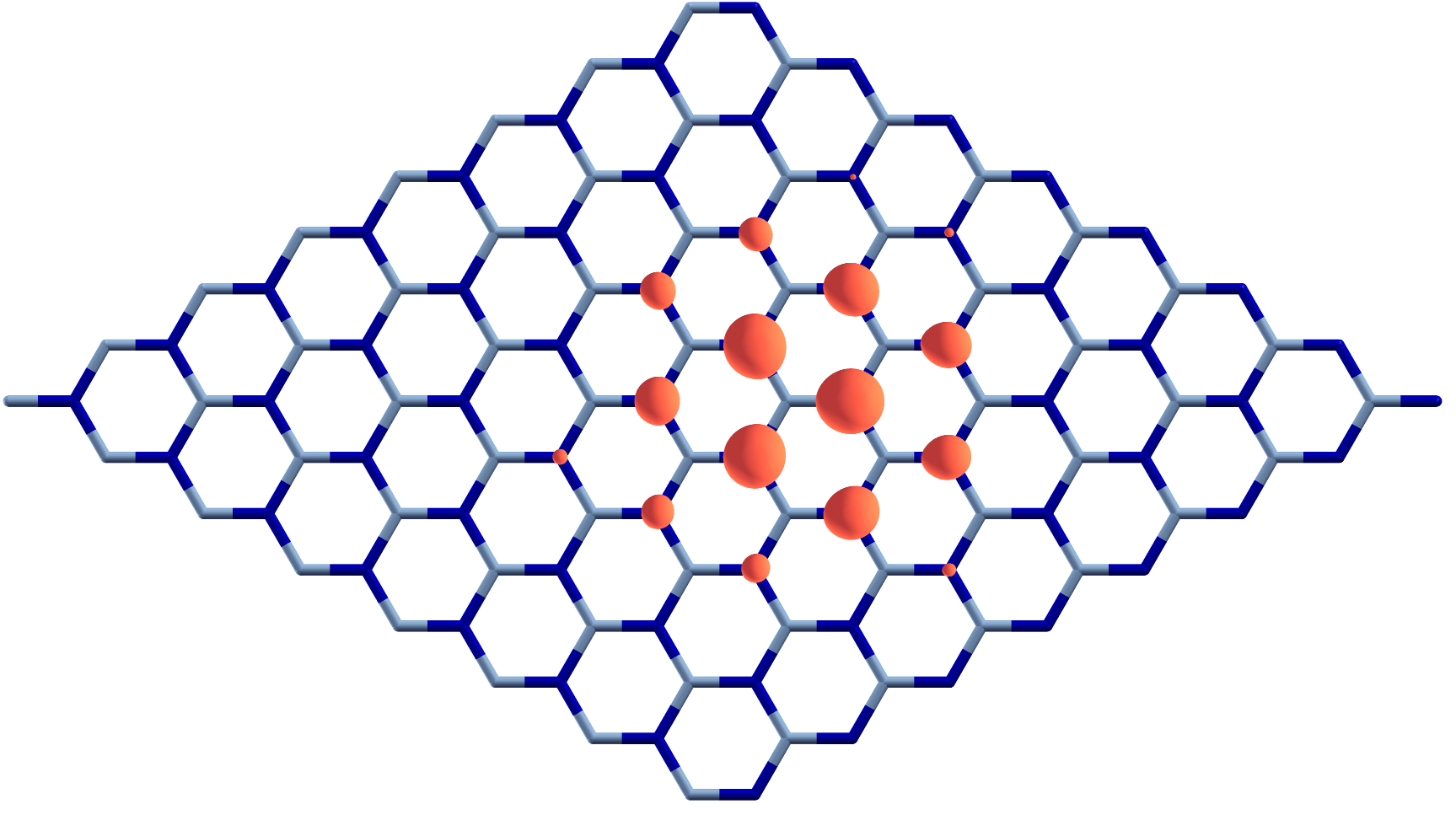} & 2\includegraphics[scale=0.1]{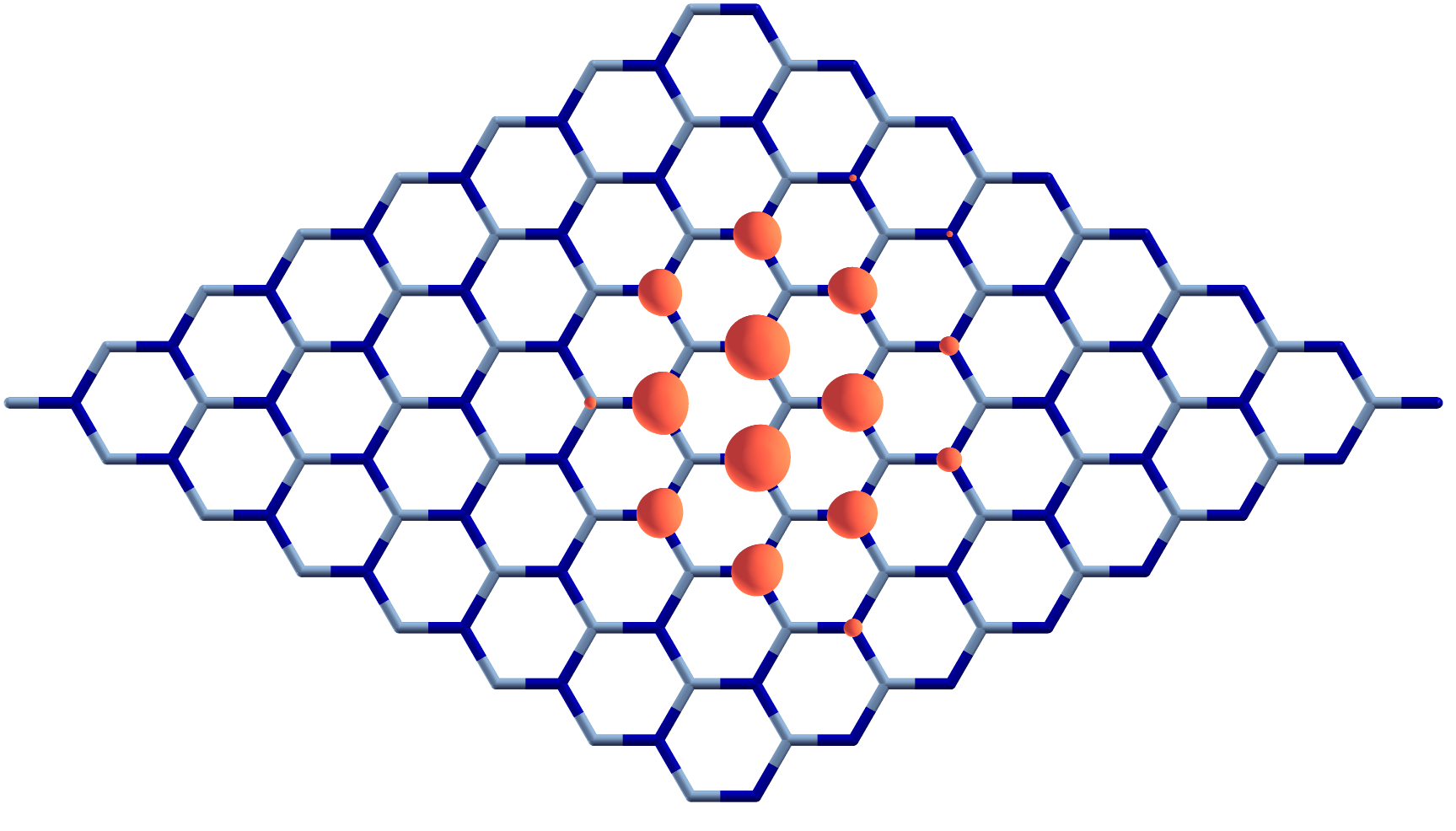}\\
 3\includegraphics[scale=0.1]{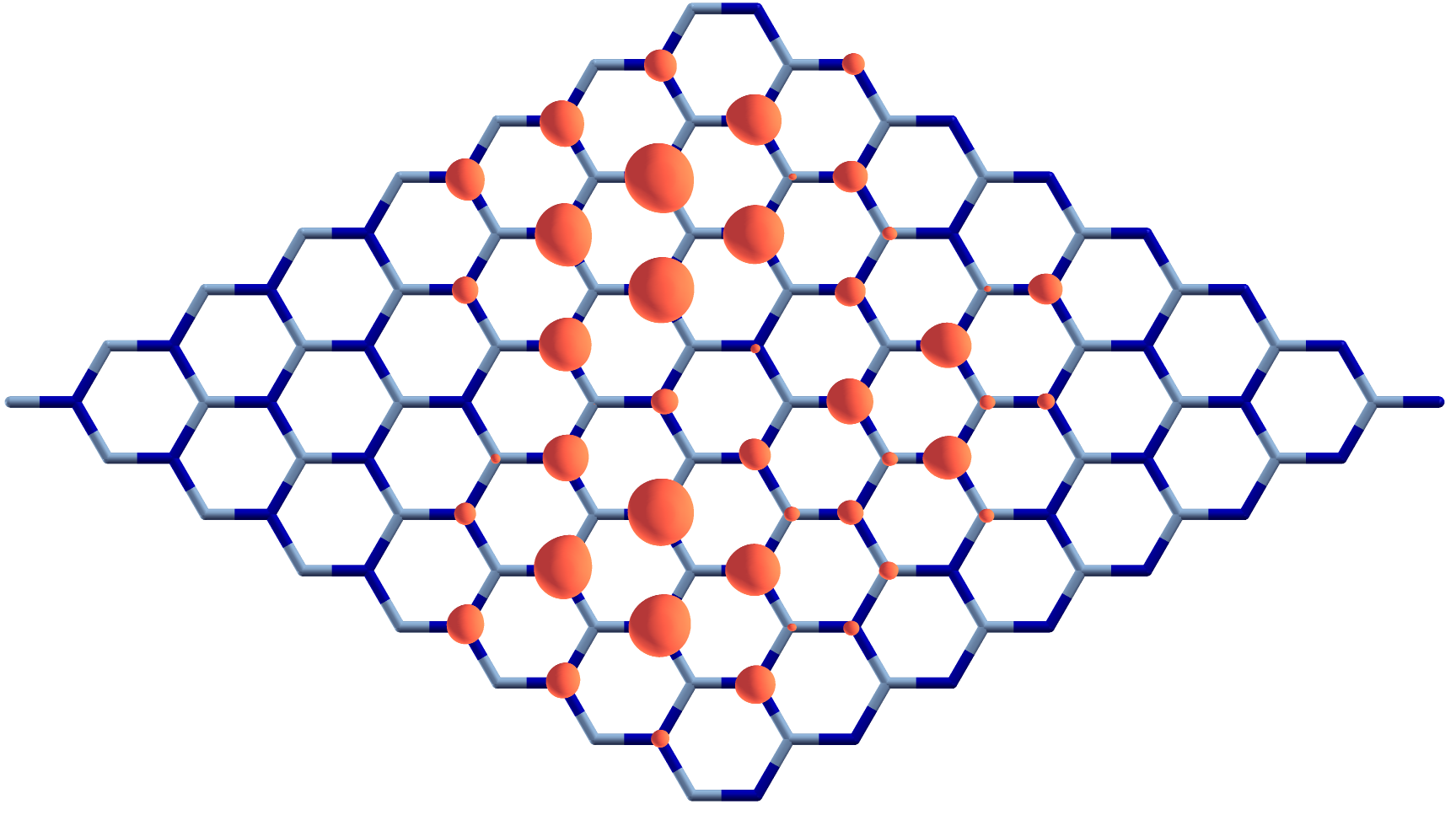} & 4\includegraphics[scale=0.1]{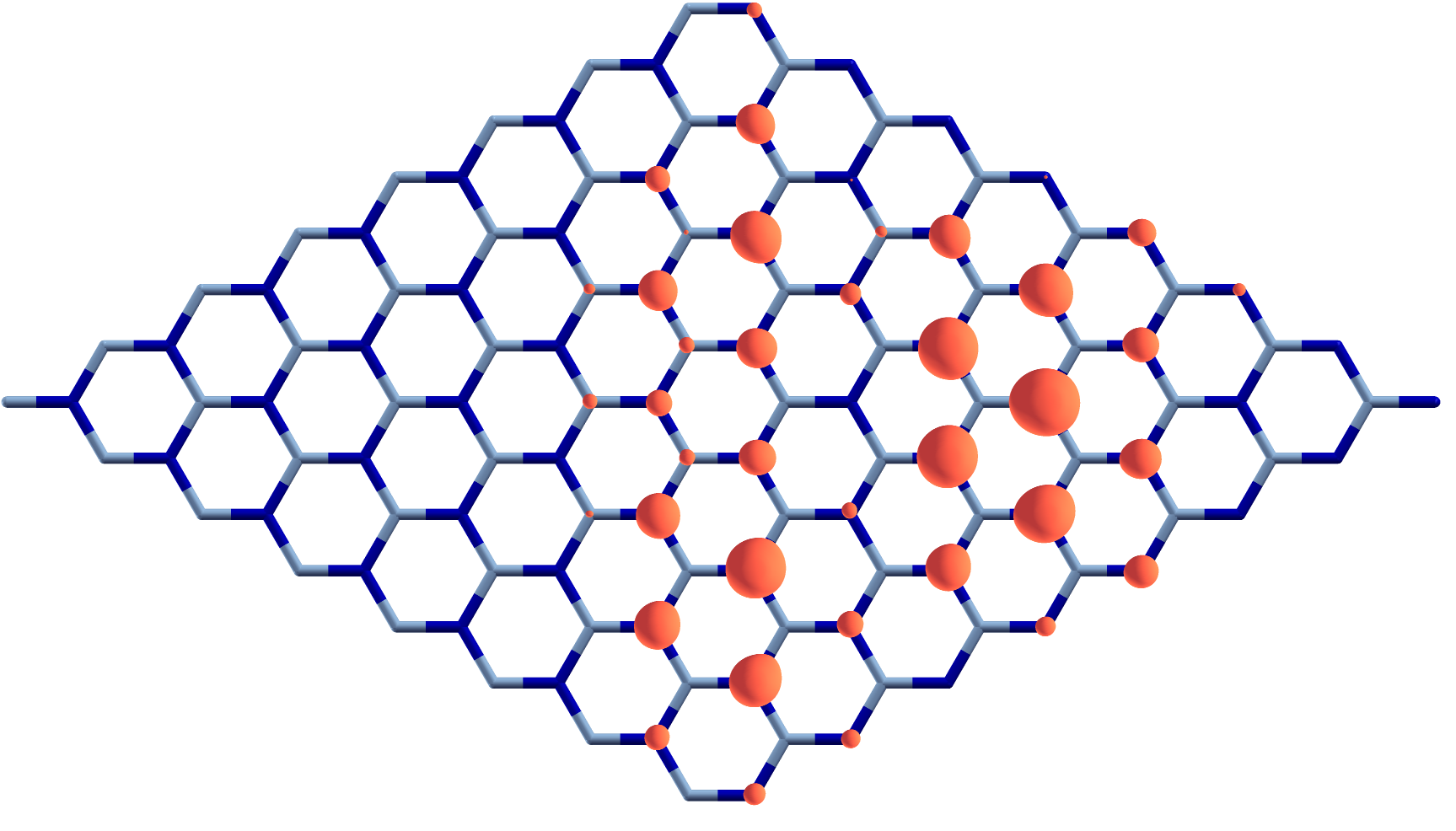} \\
 5\includegraphics[scale=0.1]{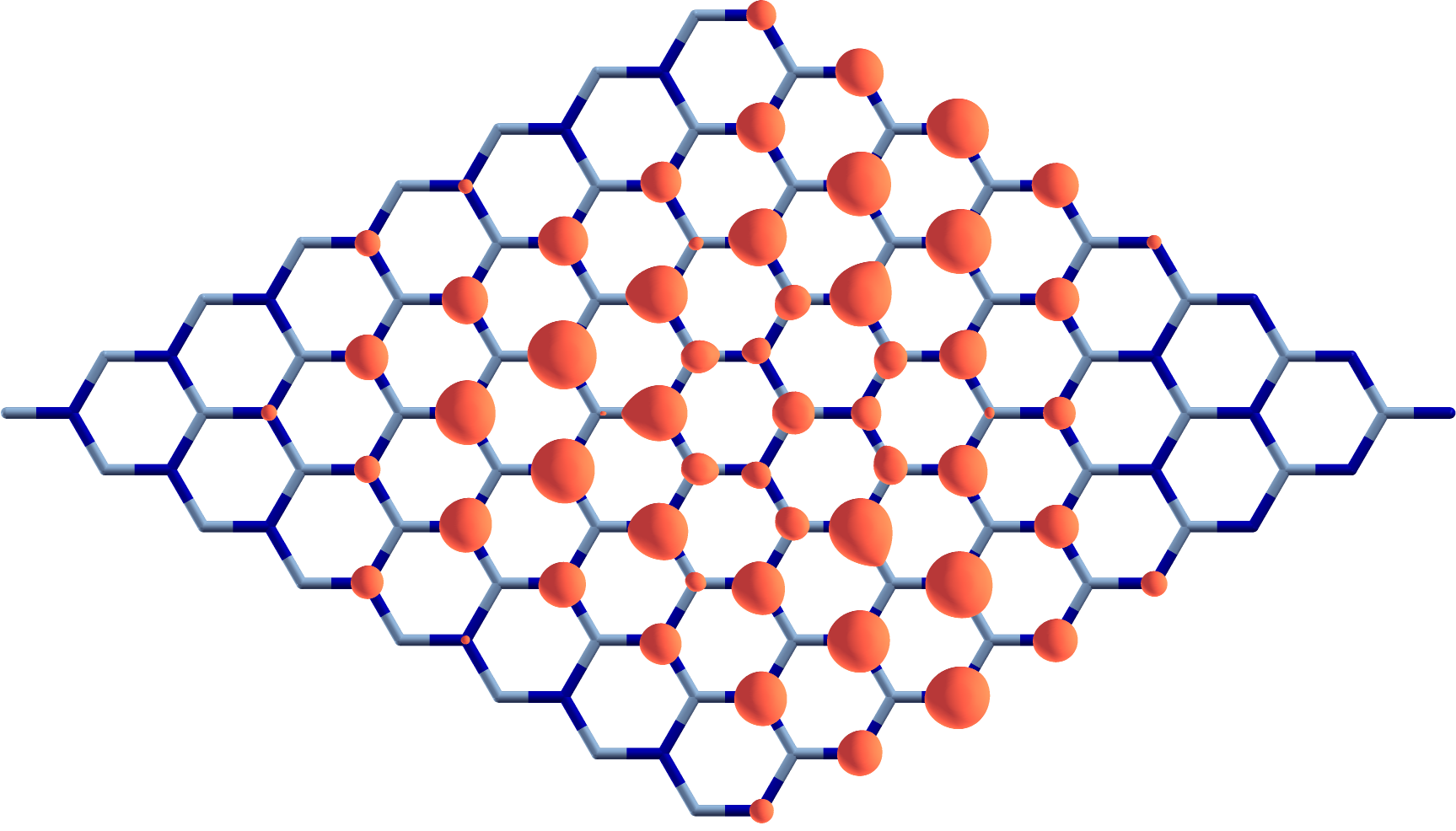} & 6\includegraphics[scale=0.1]{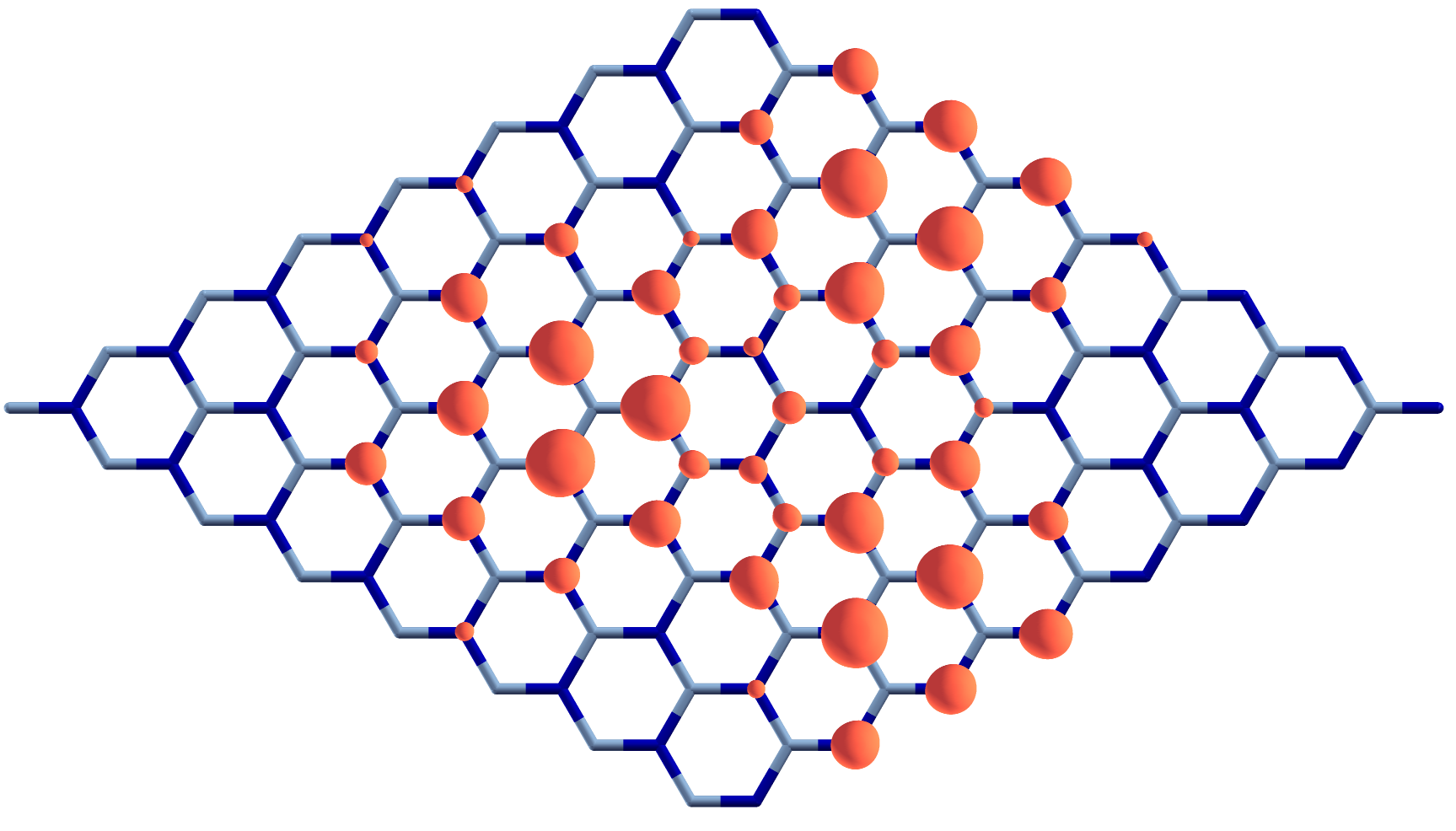}\\
 7\includegraphics[scale=0.1]{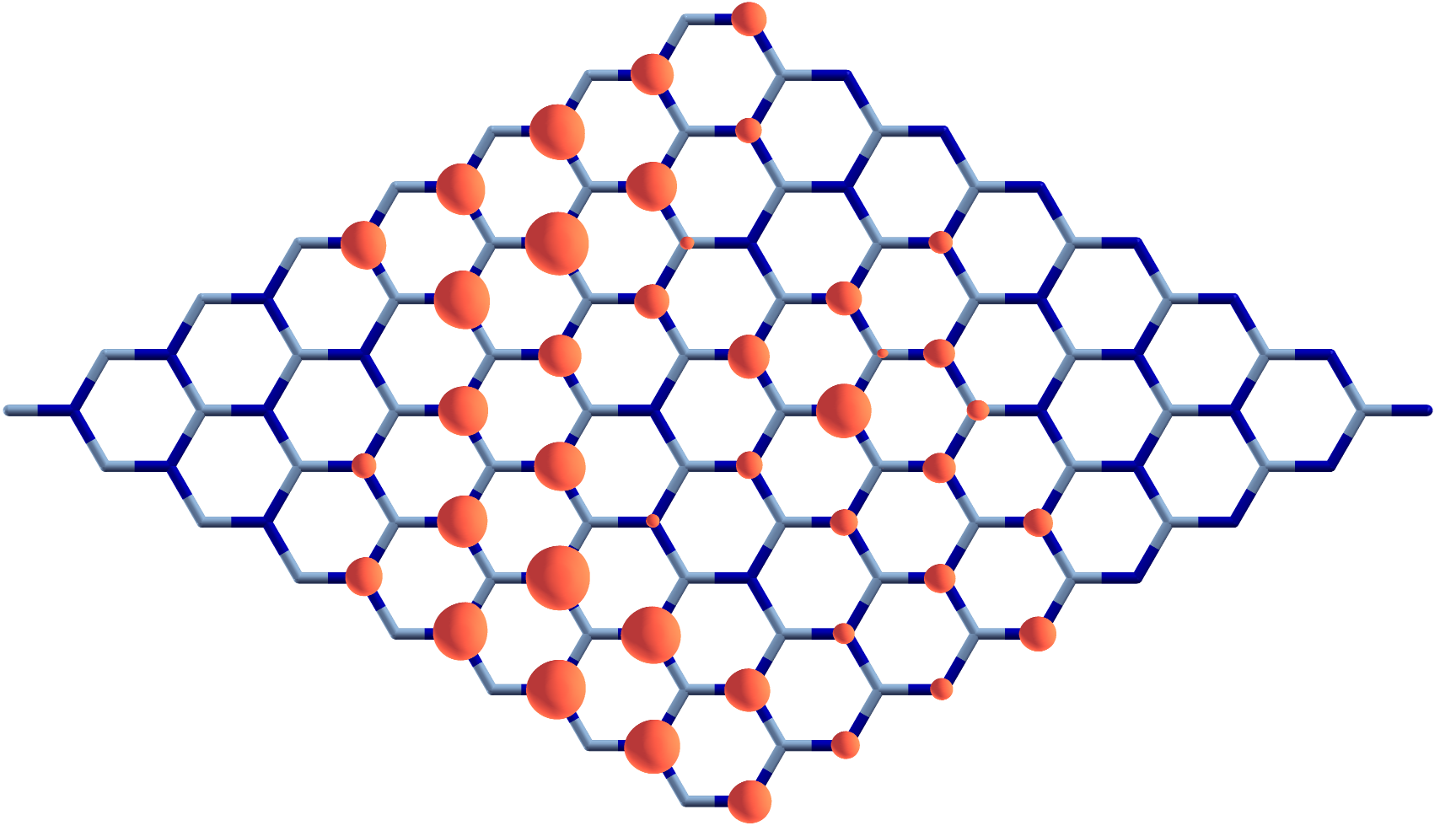} & 8\includegraphics[scale=0.1]{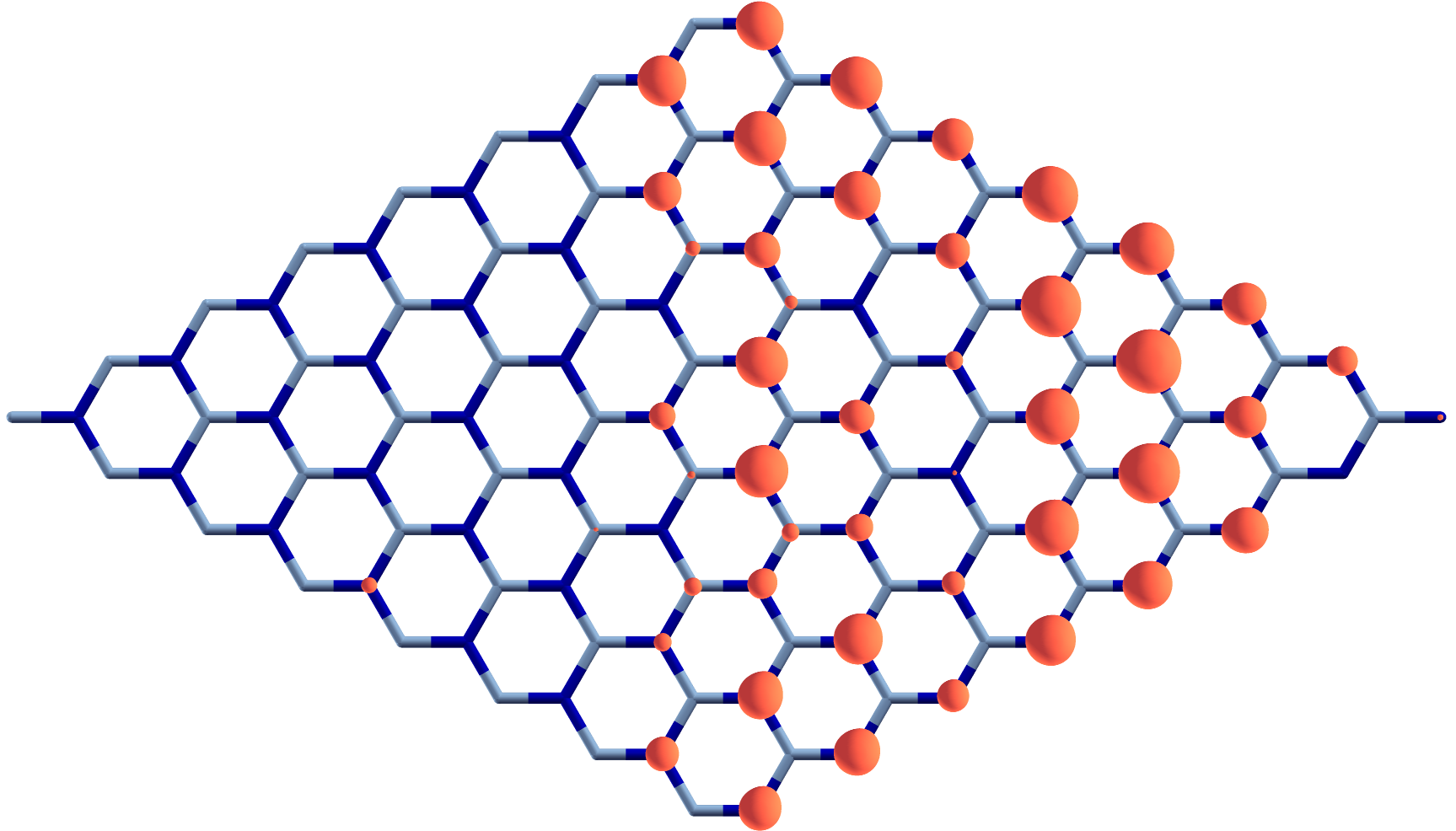}
 \end{tabular}
\caption{(Color online) Probability density $\left| \phi\left(\mathbf{r}_e, \mathbf{r}_h\right)\right|^2$ for the exciton states 1 to 8.
The hole is localized slightly above the nitrogen atom (light color) at the centre of the lattice.}
 \label{fig:prob_density}
\end{figure*}

We also calculated the eigenvalues of the two particle states.
Figure \ref{EX_energies} shows the energies of the 8 lowest energy excitonic states. From now we label each state by the corresponding energy in an ascending order.
The pairs of states (1,2), (3,4), and (7,8) are degenerate.
States 1 and 2 are the degenerated ground state.
We plot the probability density $\left| \phi\left(\mathbf{r}_e, \mathbf{r}_h\right)\right|^2$ obtained from the BSE for these eight excitonic states in
 Fig. \ref{fig:prob_density}.
These plots show the probability to find an electron at position $\mathbf{r}_e$ if the hole is located at $\mathbf{r}_h$.
We set the hole localized slightly above the nitrogen atom.
The results were calculated using a coarse grid of $12 \times 12 \times 1$  $\mathbf{k}$-points and a BSE interpolation of $72 \times 72 \times 1$  $\mathbf{k}$-points.
It can be noticed the complementarity of the degenerate states.
For instance, if one adds the probability density of states 3 and 4, the symmetry of the lattice is recovered.
And the same can be seen for the other degenerate states.
The work of Ref. \onlinecite{PhysRevB.94.125303} has also studied the excitonic states.
Their results are in good agreement with the ones obtained from this work.

\section{BSE in the equation of motion formalism and the Elliot formula}
\label{sec:Elliot}

In this section we will follow the approach of the equation of motion derived in Ref. \onlinecite{Chaves2017} and detailed in the Appendix \ref{app:eq_of_mot}.
The formalism is grounded on the calculation of the expected value of the polarization operator $\hat P(t)$ after we introduce
an external electric field of intensity ${\cal E}_0$ and frequency $\omega$ that couples with the electron gas in the 2D material.
The optical conductivity and other properties can be obtained from the macroscopic relations.
The starting point of our model is an effective Dirac hamiltonian,\cite{ribeiro2011stability}
that can be obtained from a power series expansion of the tight-binding hamiltonian.
The electron-electron interaction for a 2D material is given by the Keldysh potential.\cite{Cudazzo2011}
This effective model only considers the top valence band and the bottom conductance band.

From the equation of motion we derive the following BSE:
\be
\left(\omega -\tilde{\omega}_{\lambda\mathbf{k}} \right) p_{\lambda}(\mathbf{k},\omega)= \left( {\cal E}_0 d_\lambda(\mathbf{k})
+{\cal B}_{\mathbf{k}\lambda}(\omega) \right) \Delta f_\mathbf{k},\label{eq:bse}
\ee
where $\lambda=\pm$, $p_{\pm}(\mathbf{k},\omega)$ is the interband transition amplitude, $\tilde{\omega}_{\lambda\mathbf{k}}$ is the transition energy renormalized by the exchange self-energy and
${\cal B}_{\mathbf{k}\lambda}(\omega) $ is a term that renormalizes the Rabi-Frequency, $d_\lambda(k)$ is the dipole matrix element and $\Delta f_\mathbf{k}$ is the occupation difference, given by the Fermi-Dirac distribution. See Appendix \ref{app:eq_of_mot} for more details.

\begin{figure}[h]
\centering
\includegraphics[scale=0.5]{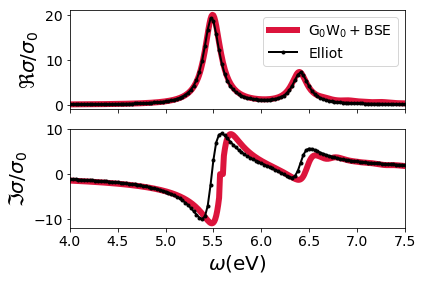}
\caption{(Color online) Fit of the Elliot formula to the $G_0W_0$+BSE result.
There is a very good agreement for the real part and a small shift in the imaginary part; the exciton linewidth used was $\gamma=0.1$~eV.
The parameters of the fitting are shown in table \ref{tab:comparison}.}
\label{fig:elliot}
\end{figure}

From the homogeneous part of Eq. (\ref{eq:bse}) we can obtain the exciton energies and the wave functions.
Using the procedure explained in Ref. \onlinecite{Chaves2017}, we can obtain the corresponding Elliot formula for the optical conductivity:
\be
\frac{\sigma(\omega)}{\sigma_0}= 4\im \hbar\omega \sum_n \frac{p_n}{\hbar\omega-E_n+\im\gamma}, \label{eq:elliot}
\ee
where $n$ labels the exciton state, $\gamma$ is the exciton linewidth, $E_n$ the exciton energy, $p_n$ the corresponding exciton weight and $\sigma_0=\frac{e^2}{4\hbar}$.
Fig. \ref{fig:elliot} shows that the $G_0W_0$+BSE described in section \ref{sec:BSE} fits well to the Elliot formula,
with a very good agreement in the real part and a small shift in the imaginary part.
The energies and weights of the fit for the $G_0W_0$+BSE and the equation of motion method are compared in table \ref{tab:comparison}.
 We use the parameters from Ref. \onlinecite{ribeiro2011stability}: $a_0=2.51$~\AA, $t_0=2.33$ eV
$\hbar v_F=\frac{\sqrt{3}}{2}  t_0 a_0$, $2m v_F^2=3.92$ eV.
The Keldysh potential parameter $r_0$ was calculated in Ref. \onlinecite{PhysRevB.94.125303} to be $r_0=10$~\AA.
We can see a excellent agreement between the exciton energies of both methods.
The difference in the weights $p_n$ can be explained by the oversimplification of the Dirac hamiltonian used for the Elliot formula
and consequently the less accurate dipole matrix elements that enter their calculation.

\begin{table}[b]
\centering
\caption{Comparison of the Elliot formula parameters used in the $G_0W_0$+BSE calculation and the equation of motion approach.
The spin and valley degeneracy is already included in the weight. } \label{tab:comparison}
\begin{ruledtabular}
\begin{tabular}{@{}lccccr@{}}
               &   $E_1$(eV)  & $p_1$    & $E_2$(eV) &   $p_2$  \\ \midrule
$G_0W_0$+BSE   &   5.48       & 0.088    &     6.41  &   0.027  \\
Eq. of Motion  &   5.52       & 0.354    &     6.53  &   0.045  \\
\end{tabular}
\end{ruledtabular}
\end{table}

\begin{figure}[h]
\centering
\includegraphics[scale=0.5]{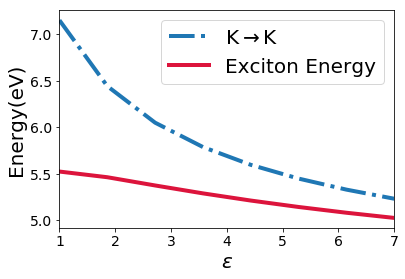}
\caption{(Color online) Exciton and K$\rightarrow$K transition energy as function of the environment dielectric constant.
We can see that the dependence of the first exciton energy is almost linear while the K$\rightarrow$K transition energy has a greater dependence on the dielectric constant. }
\label{fig:dielectric}
\end{figure}

Finally, we used the equation of motion to predict the behavior of the exciton energy and the K$\rightarrow$K transition energy
as a function of the environment dielectric constant.
The result can be seen in Fig. \ref{fig:dielectric}.
There is a strong decrease in the K$\rightarrow$K transition energy and an almost linear behavior, also decreasing,
 of the first exciton energy as the external dielectric constant increases.
This effect is simple to understand, since a large dielectric constant screens more effectively the electron-electron interaction.

\section{Exciton-Polaritons} \label{sec:Ex_pol}

In this section we discuss the exciton-polariton modes in 2D hBN.
Those modes are electromagnetic evanescent waves along the direction perpendicular to the hBN sheet. 
We assume that the hBN monolayer is cladded between two uniform, isotropic media with dielectric constants $\varepsilon_1$ and $\varepsilon_2$
and that the hBN sheet is in the $xy$-plane.
So the electromagnetic mode is evanescent in the $z$ axis and proportional to $\e^{-\kappa_i z}$ $(i=1,2)$.
The modes can be classified as transverse magnetic or transverse electric (TM/TE).

The dispersion relation for the TM mode is given by the solution given in Ref. \onlinecite{bludov2013primer}:
\be
\frac{\varepsilon_1}{\kappa_1}+\frac{\varepsilon_2}{\kappa_2}+\im \frac{\sigma(\omega)}{\varepsilon_0\omega}=0, \label{eq:tm_mode}
\ee
and for the TE mode:
\be
\kappa_1+\kappa_2-\im\omega\mu_0\sigma(\omega)=0, \label{eq:te_mode}
\ee
with $\sigma(\omega)$ the hBN optical conductivity and:
\be
\kappa_i=\sqrt{q^2-\varepsilon_i\frac{\omega^2}{c^2}},
\ee
where $q$ is the exciton-polariton in-plane wavevector and $c$ is the velocity of light in vacuum.
We shall consider the simplest case of $\varepsilon_1=\varepsilon_2=1$.
A rule of thumb is that when $\Im\,\sigma(\omega)>0$ ($\Im\,\sigma(\omega)<0$ ) TM (TE) modes are supported.

\subsection{Complex $q$ $\,\times\,$ Complex $\omega$}

First, we note that both Eqs. (\ref{eq:tm_mode}) and (\ref{eq:te_mode}) are complex.
Therefore, for a given $q$ ($\omega$) real, the solution will be a complex $\omega$ ($q$).
Each of these approaches (complex $q$ or complex $\omega$) lead to different dispersion relations for the exciton-polaritons
as discussed elsewhere.\cite{arakawa1973effect,halevi1984generalised,archambault2009surface,conforti2010dispersive,udagedara2011complex}
Both complex $q$ and complex $\omega$ approaches give the same results when an active media is used to balance the losses.\cite{udagedara2011complex}
The complex $q$ approach is suitable  when the polariton is excited in a finite region of space with a monochromatic wave, while
the complex $\omega$ approach is valid instead when the entire sample is excited by a pulsed light.\cite{archambault2009surface}

\begin{figure}
\centering
\includegraphics[scale=0.24]{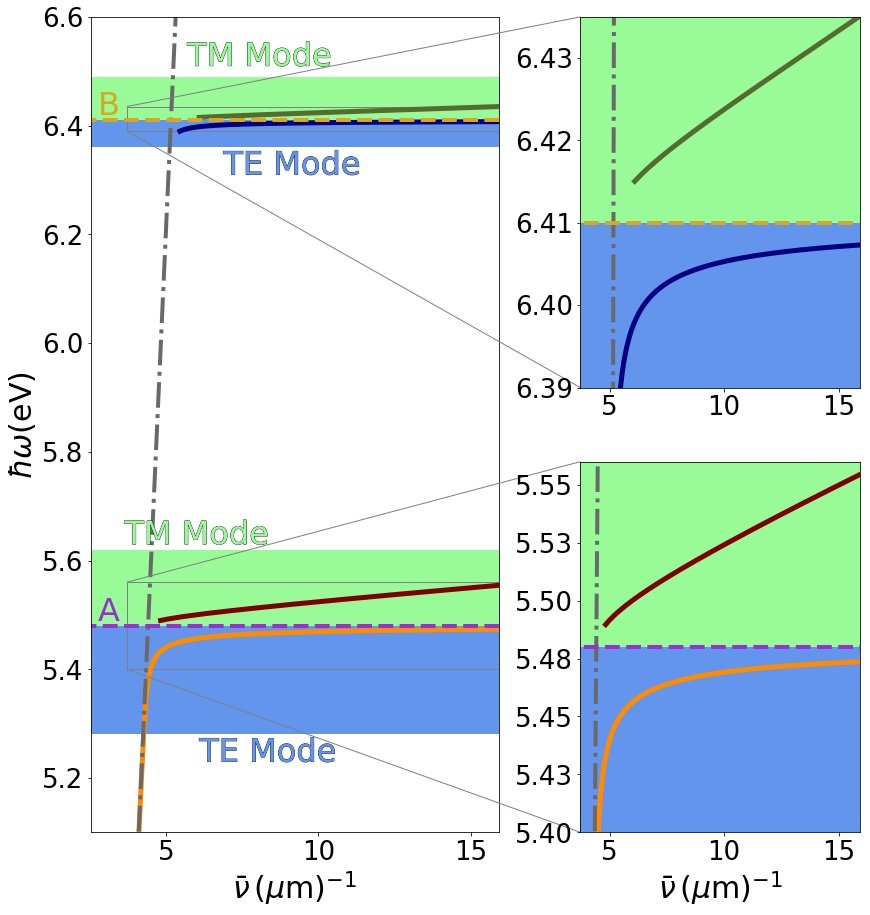}
\caption{(Color online) Exciton-polariton dispersion relation for complex frequency.
The results are given as a function of the wavenumber $\tilde{\nu}=\lambda_q^{-1}$.
The gray dashed-dot line represents the light cone in air.
In this approach, the wavenumber can reach large values for both TE and TM modes for either $A$ or $B$ exciton energies.
Detail around excitons $A$ and $B$ is shown in the right panels.}
\label{fig:dispersion_freq}
\end{figure}

The dispersion relation for both the TE and TM modes in the complex $\omega$ approach was obtained by solving Eqs. (\ref{eq:tm_mode}) and (\ref{eq:te_mode})
and using the Elliot formula (\ref{eq:elliot}) with the parameters of table (\ref{tab:comparison}) for the G$_0$W$_0$+BSE calculation and a damping of $\gamma=0.1$~eV.
The result is shown in Fig. \ref{fig:dispersion_freq}, where $A$ and $B$ denote the first two excitonic energies.
Both TE and TM modes can have a large localization (high $\kappa_i$ or $q$) in this case.
The TE mode has a flat dispersion relation that approaches the exciton energy as $q$ goes to infinity.
As expected, the TM mode has a higher frequency than the exciton energy while the TE mode has a lower frequency.
We point out that, and contrary to graphene, the TE mode presents a high degree of localization.

In the complex $\omega$ approach both excitons $A$ and $B$ support polaritons.
This can be understood by examining Eq. (\ref{eq:elliot}).
As $\hbar\omega$ approaches $E_n-\im\gamma$, the corresponding contribution to the optical conductivity diverges.
This quantity can be infinitely negative or positive depending on the real part of the frequency approaching $E_n$ from the right or the left,
supporting TM and TE modes respectively.
Fig. \ref{fig:dispersion_freq} also shows that the electrostatic limit $q\gg\omega/c$ is approached near both exciton energies.
In that limit the lifetime $\tau$ of the TM exciton-polariton $\tau=-1/\Im\,\omega$ can be calculated from (see Appendix \ref{app:exc_pol_static}):
\be
\tau^{-1}= \dfrac{\gamma}{\hbar}+\dfrac{1}{\hbar}\dfrac{p_n \Im\,[b_n]}{ \left|\dfrac{ (\varepsilon_1+\varepsilon_2)b_n  }{4\pi\alpha c q}  +1\right|^2}, \label{eq:lifetime}
\ee
where $\alpha$ is the fine-structure constant and $b_n$ is the contribution that arises from the background conductivity
provenient from interband transitions and other excitonic states.
For a negligible background $b_n\approx0$, the exciton-polariton lifetime is proportional to the inverse of the exciton linewidth $\gamma$.

\begin{figure}
\centering
\includegraphics[scale=0.4]{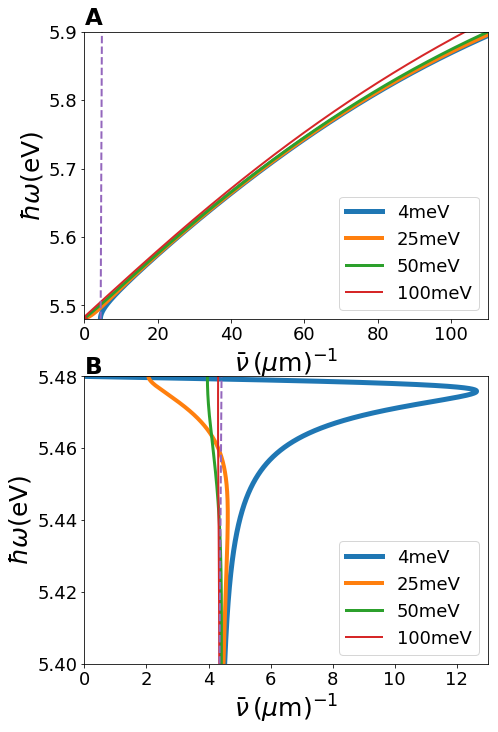}
\caption{(Color online) Exciton-polariton dispersion relation in the complex wavenumber approach.
Panel A (B) shows the TM (TE) mode.
The TM mode has a dispersion almost insensitive to the relaxation rate while the TE mode changes significantly:
the wavenumber is close to the free-light one and only for $\gamma=4$~meV there is a different behavior.}
\label{fig:dispersion_q}
\end{figure}

Next we shall consider the case of complex $q$.
There will be then a simple relation to obtain $q$ for a given frequency (assuming $\varepsilon_i=1$):
\be
c^2q^2=\omega^2+c^2\kappa^2_{\alpha}(\omega),
\ee
with $\alpha=$TM/TE and from Eqs. (\ref{eq:tm_mode}) and (\ref{eq:te_mode}) we have:
\begin{subequations}
\bea
\kappa_\mathrm{TE}(\omega)=\im\frac{\varepsilon_0\omega}{2\sigma(\omega)},
\\
\kappa_\mathrm{TM}(\omega)=\im\frac{\omega\mu_0\sigma(\omega)}{2},
\eea
\end{subequations}
The condition for the existence of polaritons is $\Re\,\kappa_\alpha>0$.
These equations allowed us to calculate the dispersion relation shown in Fig. \ref{fig:dispersion_q} for several values of the damping constant $\gamma$.
The dependence of the $\gamma$ parameter of excitons was studied for WS$_2$ in Ref. \cite{cadiz2017excitonic} as function of temperature,
showing that the linewidth decreases as the temperature decreases.
From Fig. \ref{fig:dispersion_q} we can see that the TE mode is strongly supressed except when  the damping has the very low value of $4$~meV,
close to the intrinsic line-width.
The opposite happens for the TM mode, for which the dispersion relation is almost insensitive to the damping $\gamma$.

\begin{figure}[t]
\centering
\includegraphics[scale=0.4]{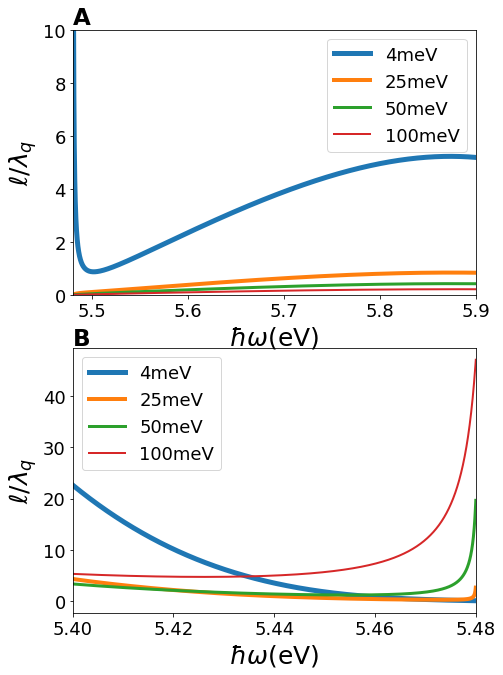}
\caption{(Color online) Exciton-polariton propagation ratio.
Panel A (B) shows the TM (TE) mode.
The propagation rate of the TM mode is very low except for $\gamma=4$~meV.
The peak at $\omega=5.48$ corresponds to the propagation of radiation.
As can be seen in Fig. \ref{fig:dispersion_q}, the wavenumber tends to the free-light wavenumber.
The same result appears in the propagation rate for the TE modes: except for $\gamma=4$~meV, all other modes correspond to poorly confined modes
(see Fig. \ref{fig:confinement} also).
For $\gamma=4$ meV and the TE mode, the propagation rate decreases with the increasing frequency.}
\label{fig:propagation}
\end{figure}

\begin{figure}[t]
\centering
\includegraphics[scale=0.35]{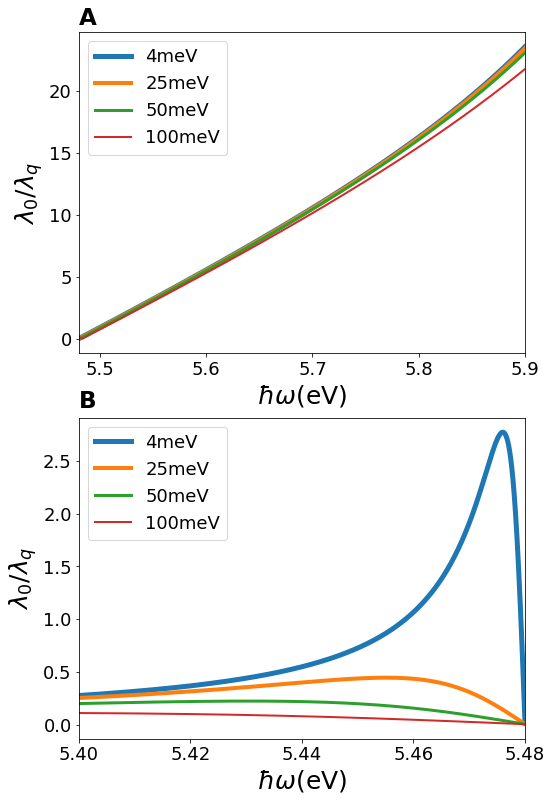}
\caption{(Color online) Exciton-polariton confinement ratio.
Panel A (B) shows the TM (TE) mode.
The confinement of the TM mode increases with the frequency and has a small dependence with the relaxation rate $\gamma$.
The TE modes for the higher values of $\gamma$ are poorly confined.
For the value $\gamma=4$ meV we have a peak in the confinement below the exciton energy.}
\label{fig:confinement}
\end{figure}

An important figure of merit is the ratio of the propagation length $\ell={\Im\, q}^{-1} $ to the exciton wavelength $\lambda_q=2\pi/\Re\,{q}$
, as it indicates if a polariton can propagate before extinction,
that is shown in Fig. \ref{fig:propagation} for several values of $\gamma$.
The TM mode is highly supressed except for the very low $\gamma=4$~meV,
while the TE mode has higher propagation rate and two different qualitative behaviors.
For larger $\gamma$, the propagation rate increases with the frequency while the opposite happens for $\gamma=4$~meV.
A better understanding of this behavior can be achieved if we consider the confinement ratio $\lambda_0/\lambda_q$,
with $\lambda_0$ being the wavelength of the free-radiation (see Figure \ref{fig:confinement}).
The confinement of the TM modes increases with increasing frequency and have a negligible $\gamma$ dependence.
On the other hand, the TE modes are poorly confined, with the confinement going to zero faster with increasing $\gamma$.
This explains the large propagation rate in this case: the poorly confined field is essentially attenuated free radiation, i.e., there are no more excitons
being excited, but the radiation field is attenuated by the material free charges.

The overall conclusion is that 2D hBN is a good platform for exciton-polaritons when we consider the complex $\omega$ approach for both TM and TE modes.
In the complex $q$ approach, the results show that exciton-polariton can be observed only for $\gamma=4$~meV.

\subsection{UV radiation mirror} \label{sec:mirror}

\begin{figure}[h]
\centering
\includegraphics[scale=0.35]{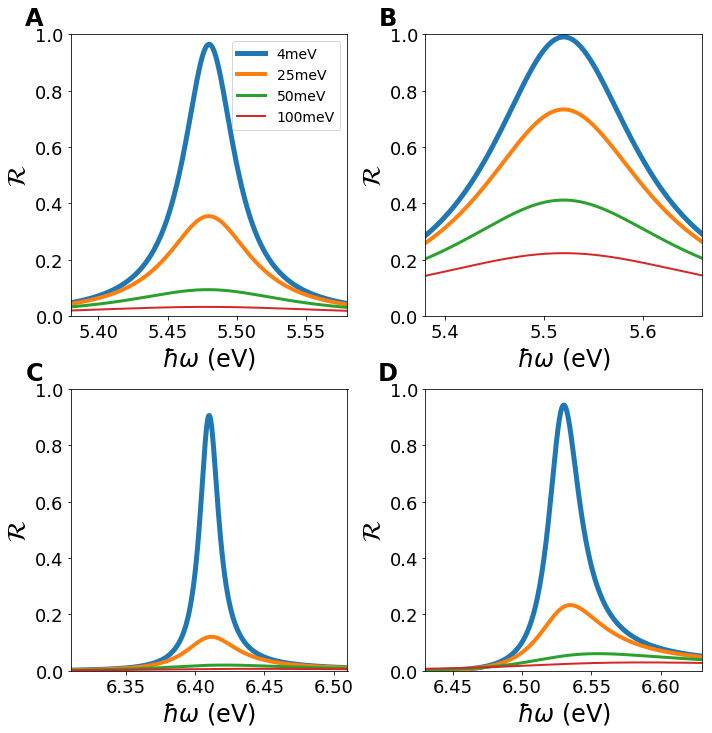}
\caption{(Color online) Reflection coefficient for monolayer hBN and different values of $\hbar\gamma$ with the parameters from table \ref{tab:comparison}. Panel (a) and (c)
shows the A and B excitons, respectively, with the G$_0$W$_0$ parameters, while the panels (b) and (d) shows the result from the equation of motion formalism. As the equation of motion
formalism predicts higher excitonic weights, in this case we have broader reflectance peaks around the excitons energies in comparison with the G$_0$W$_0$ result.}
\label{fig:mirror}
\end{figure}

It was pointed out recently that excitons in MoSe$_2$ can lead to very high reflection of electromagnetic radiation \cite{back2018realization,scuri2018large}.
In this section we show that the same occurs with hBN, but in a different spectral range.
We consider a free-standing hBN monolayer.
In this case the reflection is given by:\cite{Goncalves2016}
\be
{\cal R}= \left|\frac{\pi\alpha f(\omega)}{2+\pi\alpha f(\omega)} \right|^2,
\ee
where $f(\omega)=\sigma(\omega)/\sigma_0$ , $\alpha\approx137^{-1}$ is the fine structure constant and $\sigma_0=e^2/4\hbar$.
Fig. \ref{fig:mirror} shows that the reflection can reach almost 100$\%$ for the value $\gamma=4$ meV at the $A$ exciton energy.
This is a consequence of the very high weights for hBN that appears in the Elliot formula (see table \ref{tab:comparison}).
We emphasize that those results are for a free-standing hBN sheet.
The $\gamma$ value can be controlled by the temperature as discussed in the sections before.
As shown in Fig. \ref{fig:dielectric}, the exciton energy and therefore the reflection peak can be controlled by varying the external dielectric constant.

\section{Conclusion}
\label{sec:conclusion}

We calculated the band structure of 2D hexagonal boron nitride using DFT and the $G_0W_0$ approximation.
Then the Bethe-Salpeter equation was used to determine the excitonic energies of hBN.
We determined the values of the band gap, optical gap, excitonic binding energies using a first principles approach.
The results are in very good agreement with the ones obtained using a very different approach, namelly the equation of motion formalism and the Elliot formula,
which are also presented in this paper.
This latter formalism allowed us to study the optical properties for both the TM and TE modes.
Our results show that 2D hBN is a good candidate to polaritonics in the UV range.
We also show that a single layer h-BN can act as an almost perfect mirror for ultraviolet electromagnetic radiation.

\section*{Acknowledgments}

R.M.R. and N.M.R.P. acknowledge support from the European Commission through
the project ``Graphene-Driven Revolutions in ICT and Beyond" (Ref. No. 785219),
COMPETE2020, PORTUGAL2020, FEDER and the Portuguese Foundation for Science and Technology
(FCT) through project PTDC/FIS-NAN/3668/2014 and in the framework of the Strategic Financing UID/FIS/04650/2013.


%

\appendix
\section{Equation of motion formalism} \label{app:eq_of_mot}

The total hamiltonian that we consider in the equation of motion approach is $H=H_0+H_I+H_{ee}$ where we have the Dirac hamiltonian:
\be
H_0(\mathbf{k})=\hbar v_F\left(\boldsymbol{\sigma}\cdot\mathbf{k}+\sigma_3 m v_F^2\right), \label{H0_mos2}
\ee
the dipole interaction hamiltonian:
\be
\hat{H}_I(t)=-e{\cal E}(t) \hat{x},
\ee
and the electron-electron interaction:
\be
\hat H_\mathrm{ee}=  -\frac{e}{2}\int d\mathbf{r}_1 d\mathbf{r}_2\hat\psi^\dagger(\mathbf{r}_1) \hat\psi^\dagger(\mathbf{r}_2) V(\mathbf{r}_1-\mathbf{r}_2)\hat\psi(\mathbf{r}_2)\hat\psi(\mathbf{r}_1) , \label{hee_def}
\ee
where we used the field operator:
\be
\hat\psi(\mathbf{r},t)=\frac{1}{L}\sum_{\mathbf{k},\lambda} \phi_{ \lambda}(\mathbf{k})\hat{a}_{\mathbf{k}\lambda }(t)\e^{-\im\mathbf{k}\cdot\mathbf{r}}, \label{field_op}
\ee
with the eigenvector of $H_0$:
\be
\phi_{ \lambda}(\mathbf{k})=\sqrt{\frac{E_k+\lambda m }{2E_k}} \left(\begin{array}{cc} 1 \\ \frac{ k_x-\im k_y}{\lambda E_k+m} \end{array} \right), \label{eig_h0_ee}
\ee
and eigenvalues:
\be
E_{k}=\sqrt{k^2+m^2}.
\ee

We note that the electron-electron interaction for charges confined in a 2D material is given by the Keldysh potential:\cite{Cudazzo2011},\cite{rodin2014excitons}
\be
V(q)=-\frac{e}{2 \varepsilon_0}\frac{1}{q(r_0q+\varepsilon_m)}, \label{Keldysh}
\ee

The expected value of the polarization operator for the 2D Dirac equation can be written as:
\be
P(\omega)= -\frac{\im g_s e}{2}\sum_{\mathbf{k}\lambda} d_{-\lambda}(\mathbf{k}) p_\lambda(\mathbf{k},\omega),\label{Phat}
\ee
$g_s=4$ takes into account the spin and valley degeneracy, $\lambda=\pm$ labels the valence ($-$) or the conduction ($+$) band.
The dipole matrix element $d_{-\lambda}(\mathbf{k})$ is:
\be
d_\lambda(\mathbf{k})=-\frac{1}{2  E_k}\left(\sin\theta+\im\frac{m}{E_k}\cos\theta\right).
\ee

The interband transition amplitude is defined as:
\be
p_{\lambda}(\mathbf{k},\omega)=\int_{-\infty}^{\infty} \frac{d\omega}{2\pi} e^{-i\omega t}\left\langle\hat{a}^\dagger_{\mathbf{k},\lambda }(t) \hat{a}_{\mathbf{k},-\lambda  }(t) \right\rangle\,.
\ee
where $\hat{a}^\dagger_{\mathbf{k},\lambda }(t) $($\hat{a}_{\mathbf{k}, \lambda  }(t)$ ) is the creation (annihilation) operator in band $\lambda$ in the Heisenberg picture.

As explained in Ref. \onlinecite{Chaves2017}, from the equation of motion for the transition amplitude we can derive the following Bethe-Salpeter Equation:
\be
\left(\omega -\tilde{\omega}_{\lambda\mathbf{k}} \right) p_{\lambda}(\mathbf{k},\omega)= \left( {\cal E}_0 d_\lambda(\mathbf{k})
+{\cal B}_{\mathbf{k}\lambda}(\omega) \right) \Delta f_\mathbf{k},\label{ap_eq_mot}
\ee
where $\tilde{\omega}_{\lambda\mathbf{k}}$ is the renormalized transition energy:
\be
\tilde{\omega}_{\lambda\mathbf{k}}=2\lambda E_k+\lambda\Sigma^{\mathrm{xc}}_{\mathbf{k},\lambda},
\ee
where the exchange self-energy is included as
\bea
\Sigma^{\mathrm{xc}}_{\mathbf{k},\lambda}=\int \frac{d\mathbf{q}}{(2\pi)^2} V(q) \Delta f_{\mathbf{k}-\mathbf{q}}\big[F_{\lambda^\prime\lambda \lambda \lambda^\prime}(\mathbf{k},\mathbf{k}-\mathbf{q})
-\nonumber \\-F_{\lambda\lambda\lambda\lambda}(\mathbf{k},\mathbf{k}-\mathbf{q}) \big], \label{exchange_self_energy_0}
\eea
where  $F_{\lambda_1 \lambda_2 \lambda_3\lambda_4}$ are defined in Eq. (\ref{F_definition_ap}).
We define $\Delta f_{\lambda k}=n_F( \lambda E_k)-n_F(-\lambda E_k)$ where $n_F$ is the Fermi-Dirac distribution
and which gives us the difference in occupation between valence and conductance bands for a vertical transition.
Finally, the integral term ${\cal B}_{\mathbf{k}\lambda}(\omega)$ is:
\bea
{\cal B}_{\mathbf{k}\lambda}(\omega)= \int \frac{d\mathbf{q}}{(2\pi)^2} V(|\mathbf{k}-\mathbf{q}|)\Big[p_{\lambda}(\mathbf{q},\omega) F_{\lambda^\prime\lambda^\prime\lambda\lambda}(\mathbf{k},\mathbf{q})+\nonumber\\
+ p_{\lambda^\prime}(\mathbf{q},\omega) F_{\lambda^\prime \lambda \lambda^\prime\lambda}(\mathbf{k},\mathbf{q})\Big]\,.\label{excitonic_rabi_ap}
\eea

The homogeneous part of equation \ref{ap_eq_mot}, obtained by setting ${\cal E}_0=0$, can be used to calculate the excitons wavefunctions and energies.
From the inhomogeneous solution of $\ref{ap_eq_mot}$, $p_\lambda(\mathbf{k},\omega)$ the macroscopic polarization $P(\omega)$
can be calculated using Eq. \ref{Phat} and from there it follows the optical conductivity, permittivity and absorbance.

The overlap of four wavefunctions is given by the $F_{\lambda_1,\lambda_2,\lambda_3,\lambda_4}(\mathbf{k_1},\mathbf{k_2})$ function:
\bea
F_{\lambda_1,\lambda_2,\lambda_3,\lambda_4}(\mathbf{k_1},\mathbf{k_2})=
\nonumber\\
={{\phi^\dagger_{\lambda_1}}}(\mathbf{k}_1)
\phi_{\lambda_2}(\mathbf{k_2})
{{\phi^\dagger_{\lambda_3}} }(\mathbf{k_2})
\phi_{\lambda_4}(\mathbf{k}_1)\,. \label{F_definition_ap}
\eea

\section{Exciton in the polariton eletrostatic limit} \label{app:exc_pol_static}

In the electrostatic limit the TM exciton-polariton equation read as:
\be
\frac{\varepsilon_1+\varepsilon_2}{q}+\im \frac{\sigma(\omega)}{\varepsilon_0\omega}=0,
\ee
with the solution:
\be
\hbar\omega(q)=E_n+\frac{M_n}{ \frac{ (\varepsilon_1+\varepsilon_2)b_n }{4\pi\alpha c q}  +1}b_n-\im \hbar\gamma, \label{an_sol},
\ee
where $b_n$ can be a complex quantity, the polariton lifetime is given by $-1/\Im\,[\omega]$, and:
\be
\Im\,[\omega]= -\gamma-\frac{1}{\hbar}\frac{M_n \Im\,[b_n]}{ \left|\frac{ (\varepsilon_1+\varepsilon_2)b_n  }{4\pi\alpha c q}  +1\right|^2}, \label{eq_lt}
\ee
from (\ref{an_sol}) we can see that excitons-polaritons always exist in TMD's systems.
For the parameters considered, $ \Re\,\left[ \frac{M_n}{ \frac{ (\varepsilon_1+\varepsilon_2)b_n }{4\pi\alpha c q}  +1}b_n\right]>0$,
so the exciton-polariton will always exists for energies higher than the exciton energy.
This term also defines the exciton-polariton bandwidth, for $q\rightarrow\infty$:
\be
\Re\,\left[\hbar\omega(q\rightarrow\infty)\right]=E_n+M_n\Re\,\left[b_n\right]. \label{eq_bw}
\ee

\end{document}